# VI$_3$ - a new layered ferromagnetic semiconductor

*Tai Kong[1], Karoline Stolze[1], Erik I. Timmons[2], Jing Tao[3], Danrui Ni[1], Shu Guo[1], Zoë Yang[1] Ruslan Prozorov[2], and R.J. Cava[1]*

[1] *Department of Chemistry, Princeton University, Princeton, NJ 08544, USA*

[2] *Department of Physics and Astronomy, Iowa State University, Ames, IA 50014, USA*

[3] *Department of Physics, Brookhaven National Laboratory, Upton, New York 11973, USA*

**Abstract**

Two-dimensional (2D) materials are promising candidates for next-generation electronic devices. In this regime, insulating 2D ferromagnets, which remain rare, are of special importance due to their potential for enabling new device architectures. Here we report the discovery of ferromagnetism in a layered van der Waals semiconductor, VI$_3$, which is based on honeycomb vanadium layers separated by an iodine-iodine van der Waals gap. It has a BiI$_3$-type structure (*R*-3, No.148) at room temperature, and our experimental evidence suggests that it may undergo a subtle structural phase transition at 78 K. VI$_3$ becomes ferromagnetic at 49 K, below which magneto-optical Kerr effect imaging clearly shows ferromagnetic domains, which can be manipulated by the applied external magnetic field. The optical band gap determined by reflectance measurements is 0.6 eV, and the material is highly resistive.

# 1. Introduction

2D materials have been intensively studied for the purpose of next-generation device fabrication. Recently, several layered ferromagnetic insulators have aroused interest, as their ferromagnetism has been shown to exist down to the single-layer level[1,2]. This opens up a new route toward developing 2D devices for electronic applications by adding magnetism as a possible tuning parameter; demonstrated for example, by the magnetic field tuning of tunneling resistance in $CrI_3$ [3,4]. One of the most important factors for continuing development of 2D magnetic devices is the availability of insulating or semiconducting ferromagnetic materials with elevated magnetic transition temperatures, differing band gaps, a range of in-plane lattice parameters, and different spin configurations. This need has resulted in efforts to discover new materials both experimentally and computationally[5]. The experimental discovery of new ferromagnetic layered van der Waals materials is lagging, however, and examples of real 2D van der Waals insulating ferromagnetic materials are rare. Most well-known materials in this category are Cr-based, including $Cr_2Ge_2Te_6$ and chromium trihalides such as $CrBr_3$ and $CrI_3$. Although none of the binary early transition metal dichalcogenides order ferromagnetically in their bulk form, there are reports of ferromagnetism in $VSe_2$ and $MnSe_x$ at thicknesses down to a single-layer-level[6,7].

Here we report a new layered ferromagnetic semiconductor, $VI_3$. Compared to $CrI_3$ and other Cr-based 2D ferromagnetic insulators, in which $Cr^{3+}$ has a half filled $t_{2g}$ level yielding S= 3/2, the vanadium in $VI_3$ formally has two valence electrons that half fill two of the three degenerate $t_{2g}$ electronic states in its $VI_6$ octahedra, yielding S = 1. The partial occupancy of the $t_{2g}$ level raises the possibility for weak Jahn-Teller distortions of the V-I octahedra, orbital ordering, or complications to the otherwise potentially simple superexchange magnetic interactions in its planar honeycomb-lattice. The similarity in crystal structure between $CrI_3$, with one electron in each of the triply degenerate Cr $t_{2g}$ states, and VI$_3$, with two electrons in the triply degenerate V $t_{2g}$ states, offers new possibilities for tuning the magnetism in layered electronic devices.

## 2. Results and Discussion

Transition metal halides, as simple binary systems, have been studied for decades[8]. The existence of VI$_3$ was in fact reported in 1969[9,10] but only the unit cell dimensions and space group were described, with no structural characterization or physical property determination. This compound has thus largely been ignored. In searching for new layered ferromagnetic materials, we have re-discovered this material and grown relatively large single crystals of VI$_3$. The dark, hexagonal plate single crystals are sensitive to humidity, are soft, and are very easy to cleave.

We have determined the crystal structure at 100 K for VI$_3$ via single crystal x-ray diffraction (SXRD). The material has the layered BiI$_3$ type structure (Figure. 1, space group *R*-3, No. 148), based on VI$_6$ octahedra sharing edges within each layer to form a honeycomb lattice. The honeycomb layers are then stacked along the *c*-axis in an ABC stacking sequence. This structure is often observed for transition metal tri-halides[8] such as CrCl$_3$, CrI$_3$ [11,12], VCl$_3$ and VBr$_3$[8]. In the case of VI$_3$, the normally vacant interstitial sites within the honeycomb layers appear to be partially occupied by a small number of additional vanadium atoms. The partial occupancy is at only the 4% level. Because diffraction measurements are a positional average over the whole crystal, it is possible that the apparent 4% excess vanadium in the normally vacant sites in the honeycomb layer in the VI$_3$ structure is due to the presence of stacking faults (which make the average structure appear to have atoms in normally vacant sites in the honeycomb layers even though they are not really present). Although no evidence for stacking faults, as are commonly seen in similar layered structures such as in RuCl$_3$[13], was observed in our diffraction data, we cannot rule out stacking faults as the origin of the apparent interstitial vanadium. And the absence of intralayer disorder can be inferred from the sharp transitions we have observed in the specific heat (see below). The atomic positions for $VI_3$ at 100 K can be found in Table 1, and the single crystal refinement parameters are found in the supplementary information, Tables S1 and S2.

First we address the semiconducting character of $VI_3$. Our optical absorption data (Figure 2a) indicate the presence of a bulk optical band gap of ~0.6 eV at room temperature, which is smaller than that for $CrI_3$ (1.2 eV) [14]. The measured in-plane resistivity of $VI_3$ is in the order of 10 Ω m at 360 K, with an activation-energy of 0.4 eV (Figure 2a, lower inset). These data suggest a consistent semiconducting behavior of $VI_3$ and that the small amount of conductivity in $VI_3$ is likely due to the excitation of carriers from a donor states to the conduction band. The calculated electronic structures with and without spin orbit coupling are shown in Figure 2b. As is frequently the case for materials where the electrons at the Fermi energy arise predominantly from metal atoms where electronic correlations are significant, such as vanadium, the calculated electronic structure for the non-magnetic case is metallic. This is in a clear contrast to the experimentally observed semiconducting behavior at room temperature, well above its magnetic ordering temperature. This inconsistency between experiments and band structure calculation also exists in $CrI_3$, and was attributed to a potential Mott insulating state[11]. Thus the strong correlation effect may be a necessary component for a better understanding of the electronic property of $VI_3$.

Having established that the material is semiconducting with a band gap near 0.6 e V and a high resistivity, we now turn to its magnetic characterization. The magnetic susceptibility data for $VI_3$ are shown in Figure 3(a). At high temperatures, the susceptibility follows a Curie-Weiss law, with an effective moment of ~2.6 $\mu_B$/V, which is close to the expected value for a spin-1 configuration of $V^{3+}$(~2.83 $\mu_B$/V). The Curie-Weiss temperature is 25 K, which, because it is lower than the magnetic ordering temperature of 49 K, indicates the presence of competing ferromagnetic and antiferromagnetic interactions at high temperatures.

At low temperatures, phase transitions were observed. As mentioned above, the material goes through a ferromagnetic phase transition at around 49 K. In addition, there is another, more subtle, phase transition at ~ 78 K. This higher-temperature phase transition is more clearly seen in the magnetization measurement with the magnetic field applied along the *c*-axis.

At 1.8 K, well below the magnetic ordering temperature, the magnetic anisotropy is weak yet clear. The easy axis lies along the *c*-axis, and the magnetization does not saturate in the applied fields tested, nor does it approach the expected saturation value for a spin-1 system. This is in contrast to $CrI_3$[11,15] where saturation is reached by 30 kOe for its hard axis at 2 K. The magnetization values for *H//c* and *H//ab* are different at 90 kOe, which is reproducible in all of our measured samples. The distribution of anisotropic magnetization values at 90 kOe are shown by error bars in Figure 3(b). It is possible that this difference arises because the magnetic moments are canted or have a small antiferromagnetic component. Future neutron scattering experiments would be of interest to clarify the magnetic structure of $VI_3$.

The magnetic hysteresis loop is strong and relatively sharp for $VI_3$ below its magnetic ordering temperature. For *H//c*, the coercive field is close to 10 kOe, and for *H//ab*, the coercive field is much smaller, at around 1 kOe. Both are significantly larger than those in $CrI_3$[15]. The magnetic hysteresis loops shown in Figure 3b also show a clear magnetic domain creep behavior, manifested in the sudden jumps of the magnetization in applied field. The originating effect, domain wall motion, not only happens when the magnetic field is changed, it also evolves with time in the absence of external magnetic field, suggesting a small domain-wall-energy for this material.

Figure 3c shows the ferromagnetic domain evolution at 5 K as a function of applied magnetic field. Both positive and negative field data were collected after a zero field cooling (ZFC) process. The magnetic domains are of the stripe type, which are perpendicular to larger, structural domains. The typical domain size at zero-field is about 10 microns. With increasing applied magnetic field, domain walls move and magnetic domains merge together. The domain morphology was studied in 2D thin film magnetic systems to investigate the nature of their ferromagnetism close to spin reorientation transition [16]. In the van der Waals ferromagnetic semiconductor community, studies of domain morphology in relation to its ferromagnetism is an unexplored area. Further research may reveal insights into magnetic domain formation in atomically thin 2D ferromagnets. In the current study, the appearance of magnetic domains serves as an additional clear illustration of the ferromagnetism in $VI_3$.

We now look further into the phase transitions that are observed in our magnetic susceptibility data. In Figure 4, the specific heat of $VI_3$ is plotted against temperature. It shows two phase transitions, at 78 K and 49 K. No features that can be associated with possible $VI_2$ [17] phases were observed. Both transitions are sharp in temperature. The estimated Debye temperature is around 143 K, close to that of $CrI_3$ [11]. As stated previously, we postulate that the higher temperature phase transition is a structural phase transition. This is supported by the observation of emerging domains as shown in the upper middle inset in Figure 4 with different polarized light contrast. These additional domains emerge abruptly as the temperature was lowered across the 78 K (see more details in Supplementary Information). These domains are different from the magnetic domains, because they are not sensitive to external magnetic field, consistent with purely paramagnetic-like field-dependent magnetization measured at 70 K (Figure 3 (b)). Furthermore, when fitting with the Curie-Weiss law, the effective moment in between 78 K and 49 K remains 2.6 $\mu_B$/V, which is inconsistent with a magnetically ordered state in this temperature range. Therefore it is highly plausible that these optical domains originate from the twin structures formed below a structural phase transition.

To better understand the phase transition at 78 K, we conducted electron diffraction measurements in a transmission electron microscope (TEM) at ambient temperature and 35 K. Surprisingly, when compared to the room temperature data, no clear superstructure diffraction peaks or structural distortions were observed in the TEM at 35 K (Figure 4 lower panel). The symmetry along the *c*-axis remains the same for the whole temperature range of measurement, and the *hk0* planes shown in the figure show no apparent temperature-induced differences. Although the high dynamic range of electron diffraction does allow us to say with confidence that there cannot be a superlattice present at low temperatures, a structural distortion that does not increase the unit cell volume could be present, which would be smaller than is possible to observe by our methods. In the case of transition metal dichalcogenides, the structural phase transition temperature can be changed when the sample thickness is reduced to several atomic layers[18,19]. However, in our measurement, sample thickness is of the order of 20-50 nm, which is well beyond the range of concern. These results suggest that a more productive way to study the 78 K phase transition would be by high resolution synchrotron x-ray diffraction.

**Conclusion**

In summary, we have discovered a new vanadium-based van der Waals layered, 2D ferromagnetic semiconductor, $VI_3$. It shows two phase transitions at low temperatures, one at 78 K that is likely a structural phase transition and one at 49 K, which is a ferromagnetic phase transition. The high temperature paramagnetic state of $VI_3$ follows the Curie-Weiss behavior very well for a spin-1 system, as expected by straightforward electron counting. In the ferromagnetic regime, the vanadium magnetic moment does not saturate up to 90 kOe and differs for in-plane and axial orientations. In comparison with chromium-based, van der Waals ferromagnetic semiconductors, the discovery of $VI_3$ offers more opportunities in studying the ferromagnetism in layered structures and more tuning possibilities in future 2D devices.

**Experimental methods**

VI$_3$ single crystals were synthesized by a chemical vapor transport method. A stoichiometric mixture of vanadium powder and iodine pieces was loaded in a silica tube (ID 14 mm, OD 16 mm, length 12 cm) and sealed under vacuum. The ampoule was then put into a horizontal tube furnace, with the hot end held at 400 °C and the cold end at ambient temperature. After 3-4 days of vapor transport, the ampoule was taken out of the furnace and air-quenched. Milimeter sized black plates, as shown in Figure 3(b), were obtained.

The crystal structure of VI$_3$ was determined by SXRD. The SXRD data was collected at 100 K with a Bruker D8 VENTURE diffractometer equipped with a with PHOTON CMOS detector using graphite-monochromatized Mo-K$_\alpha$ radiation ($\lambda = 0.71073$ Å). The raw data was corrected for background, polarization, and Lorentz factor using APEX3 software[20] and multi-scan absorption correction using the SADABS-2016/2 program package was applied[21]. The structure was solved with the charge flipping method[22] and subsequent difference Fourier analyses with Jana2006[23,24]. Structure refinement against $F_o^2$ was performed with Shelxl-2017/1[25,26].

Magnetization measurements were performed using a Quantum Design (QD) Physical Property Measurement System (PPMS) Dynacool equipped with a VSM option. For magnetic field along *c*-axis, single crystals were mounted in the standard plastic capsule and for field along *ab*-plane, samples were glued onto a silica sample holder with GE-varnish. Heat capacity data were also collected using a QD PPMS Dynacool. dc resistance was measured using a QD PPMS. Platinum wires were attached to samples via DuPont 4922N silver paint.

All samples were prepared in an argon glove box and quickly transferred to the instrument prior to measurement.

Kerr effect polarized light optical microscopy was used to identify the locations of magnetic domains in temperatures below Curie temperature or to indicate the lack of magnetic domains above $T_c$. In the magnetically ordered phase the permittivity is anisotropic and, depending on the angle between polarization direction of the incident light and magnetic anisotropy axis, domains of different orientations will have different optical contrast when observed through an analyzer positioned close to 90° with respect to the polarizer axis. In our setup linearly polarized incident light is perpendicular to the sample surface.

The diffuse reflectance spectrum was collected by a Cary 5000i UV-VIS-NIR spectrometer equipped with a DRA-2500 integrating sphere on powder samples and reflectance data were converted using the Kubelka-Munk method[27]. The band gap was estimated based on the relation $\alpha h\upsilon = A(h\upsilon - E_g)^n$, where A is a constant, $\alpha$ is the absorption coefficient ($cm^{-1}$), $E_g$ is the band gap and n is 0.5 for direct transition (n = 2 for indirect transition)[27].

The DFT calculations for the band structure plot were executed using the projector augmented wave method[28] and the Vienna ab initio software package (VASP)[29–31]. A standard self-consistent run was performed using a POSCAR file produced from a cif file by VESTA, and an automatically generated, gamma-centered, 11x11x11 K-mesh. The high-symmetry path for the non-self-consistent band structure calculation was obtained according to the conventions listed[32]. Pymatgen[33] was then used to parse and plot the band structure from the output of the non-self-consistent calculation.

Electron diffraction experiments were carried out using a JEOL ARM 200F transmission electron microscope that has a cold-field emission gun and double-Cs correctors. The TEM sample of $VI_3$ was prepared in a glove box in argon gas and was transferred into the electron microscope without exposure to air. The in-situ cooling experiment was performed using a Gatan liquid-helium sample holder that enables the temperature-dependent control and measurements. The electron diffraction results shown here were obtained from the same volume of a thin flake in the TEM sample during cooling and warming cycles and the results are representative according to the observations from a few thin flakes.

**Acknowledgements**

We thank M. A. Tanatar for valuable discussion regarding magneto-optical experiments. Work conducted at Princeton University was supported by the NSF-sponsored PARADIGM program centered at Cornell University, grant DMR-1539918. Work in Ames was supported by the U.S. DOE Office of Science, Basic Energy Sciences (BES), Materials Science and Engineering Division. Ames Laboratory is operated for the U.S. DOE by the Iowa State University under contract no. DE-AC02-07CH11358. Work at Brookhaven National Laboratory was funded by the DOE BES, by the Materials Sciences and Engineering Division under Contract DEAC02-98CH10886.

**Figure Captions**

**Figure 1. Crystal structure of VI$_3$.** Crystal structure of VI$_3$ viewed along the (a) *a*-axis and (b) *c*-aixs. VI$_6$ octahedra form a layered honeycomb lattice via edge-sharing within each layer and layers are stacked in an ABC sequence along the *c*-axis. Honeycomb lattice of vanadium is illustrated in (b) via yellow solid lines. Possible partial occupancy of the vanadium atoms are located in the normally vacant interstitial sites in the honeycomb array, represented by a fraction of orange color in the white atomic sites.

**Figure 2. Optical absorption, resistivity and band structure of VI$_3$**. (a) The optical absorption of VI$_3$ as a function of wavelength shows a clear drop starting from ~1500 nm. The upper inset shows $(\alpha h\upsilon)^2$ (eV/cm)$^2$ plotted against photon energy to estimate the optical band gap for a direct transition. The optical bandgap was obtained by extrapolating the linear absorption region through the red lines until it intersects with the baseline of the absorption. Indirect band gap estimation gives similar gap value. The lower inset shows the dc resistance as a function of inverse temperature in a semi-log plot. The resistivity value at 360 K is in the order of 10 Ω m. Both band structure calculations (b) without and (c) with the spin-orbit coupling show metallic behavior, in contrast to optical and resistive data. Strong correlations may need to be considered to understand the electronic behavior of VI$_3$.

**Figure 3. Magnetic properties of VI$_3$.** (a) The inverse magnetic susceptibility of VI$_3$. The upper left inset shows the low temperature anisotropic magnetization measured at 500 Oe, showing a clear ferromagnetic phase transition. The lower right inset shows zoom-in-view at the structural transition for *H//c*. (b) Field dependent magnetization of VI$_3$ at 1.8 K and 70 K. Strong magnetic hysteresis was observed for both orientations at 1.8 K whereas the field dependent magnetization measured at 70 K is consistent with a paramagnetic behavior. The left inset shows the initial magnetization curve for both orientations. It is weakly anisotropic with *c*-axis being the easy axis. The magnetization values are lower than the expected saturation value for spin-1 V$^{3+}$ up to 90 kOe. The right inset shows VI$_3$ single crystals on a millimeter grid paper. (c) Magneto-optical characterization of the ferromagnetic domains in VI$_3$ at 5 K. Both measurements were conducted after a ZFC. Magnetic domain stripes are perpendicular to structural domains. Typical width of a magnetic domain is ~ 10 μm. With increasing magnetic field, the domains quickly merge together.

**Figure 4. Phase transitions of VI$_3$**. The temperature-dependent specific heat of VI$_3$ is shown in the central figure. Two strong peaks appear at 78 K and 49 K. The upper panels show magneto-optical data measured in each phase. Structural domains emerge abruptly when VI$_3$ is cooled below 78 K. The magnetic domains appear below 49 K within the existing structural domains. Lower panels show the electron diffraction characterization of VI$_3$ at 35 K and 295 K. No additional superstructure diffraction peaks were observed in the data well below 78 K.

.

**Figure 1**

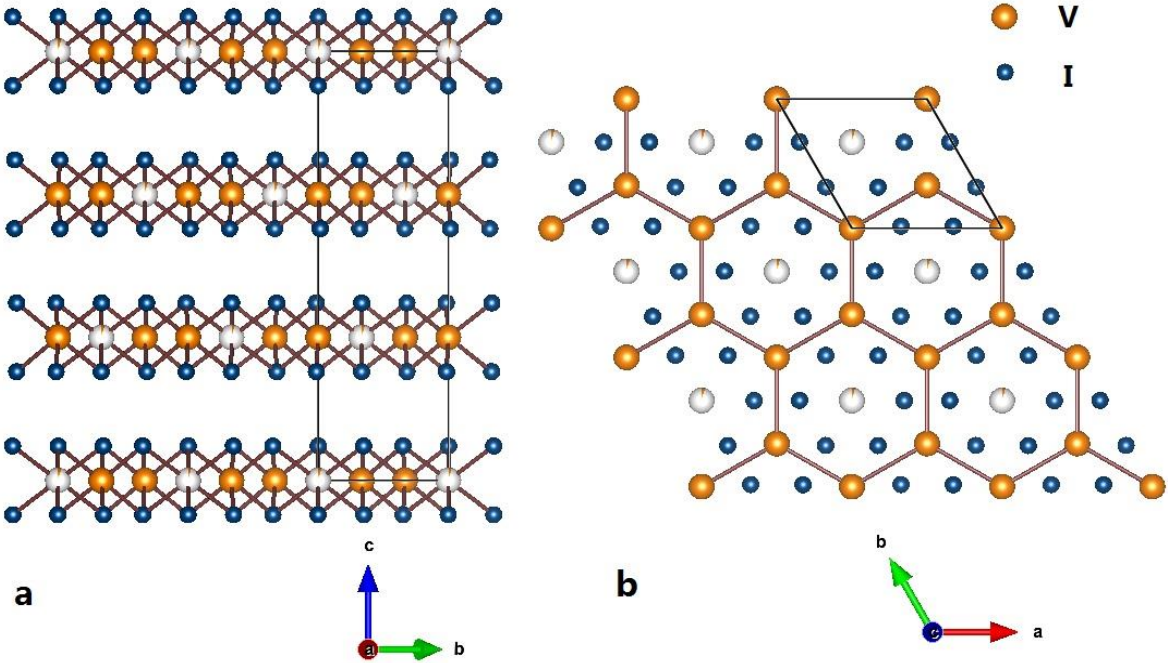

**Figure 2**

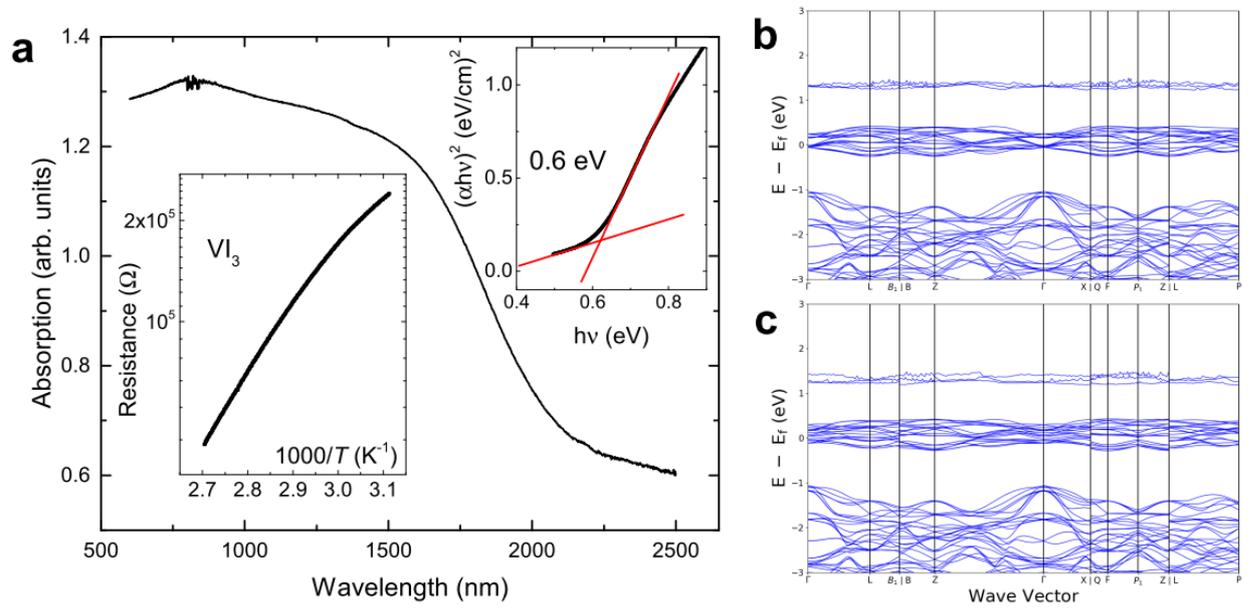

**Figure 3**

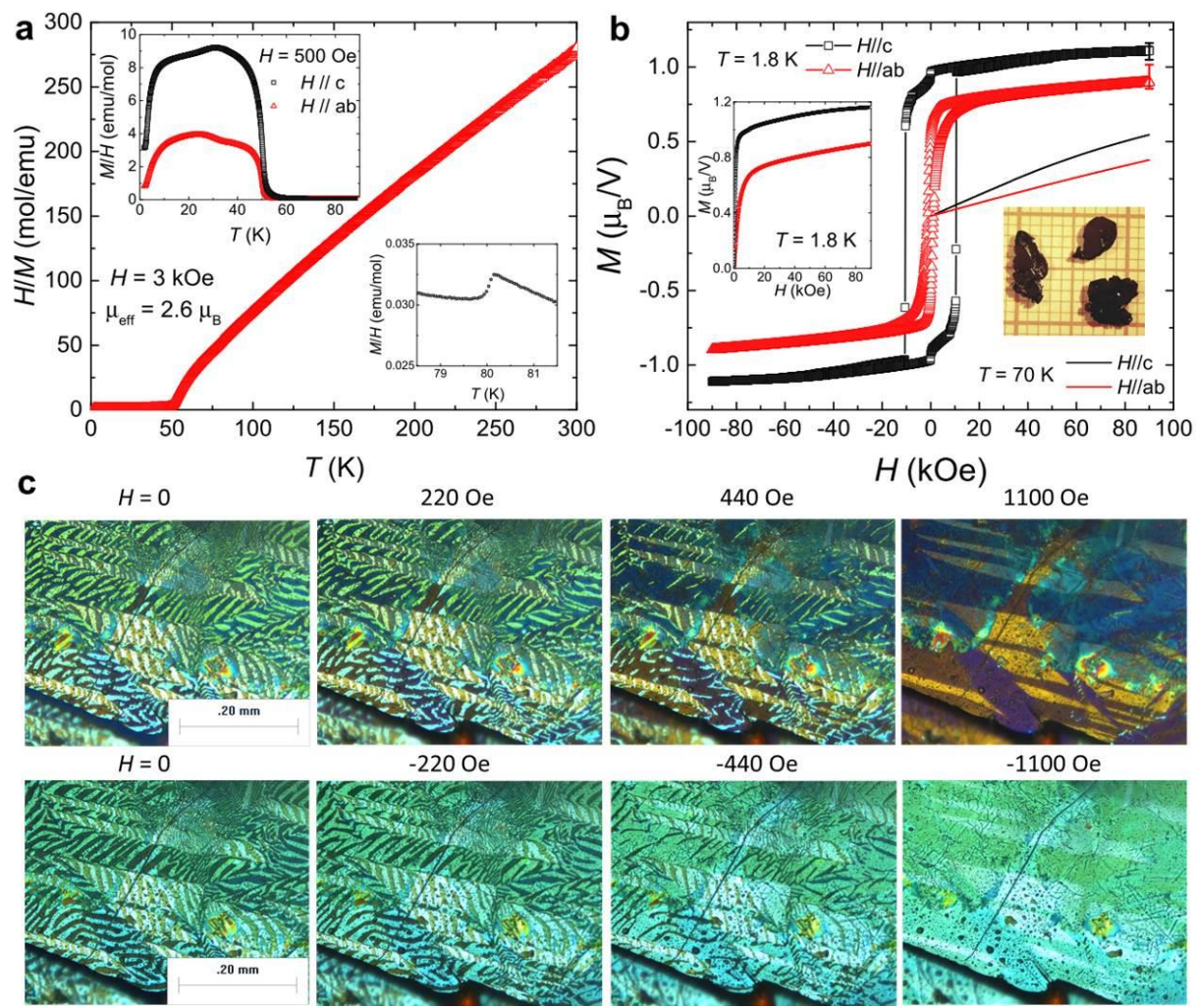

**Figure 4**

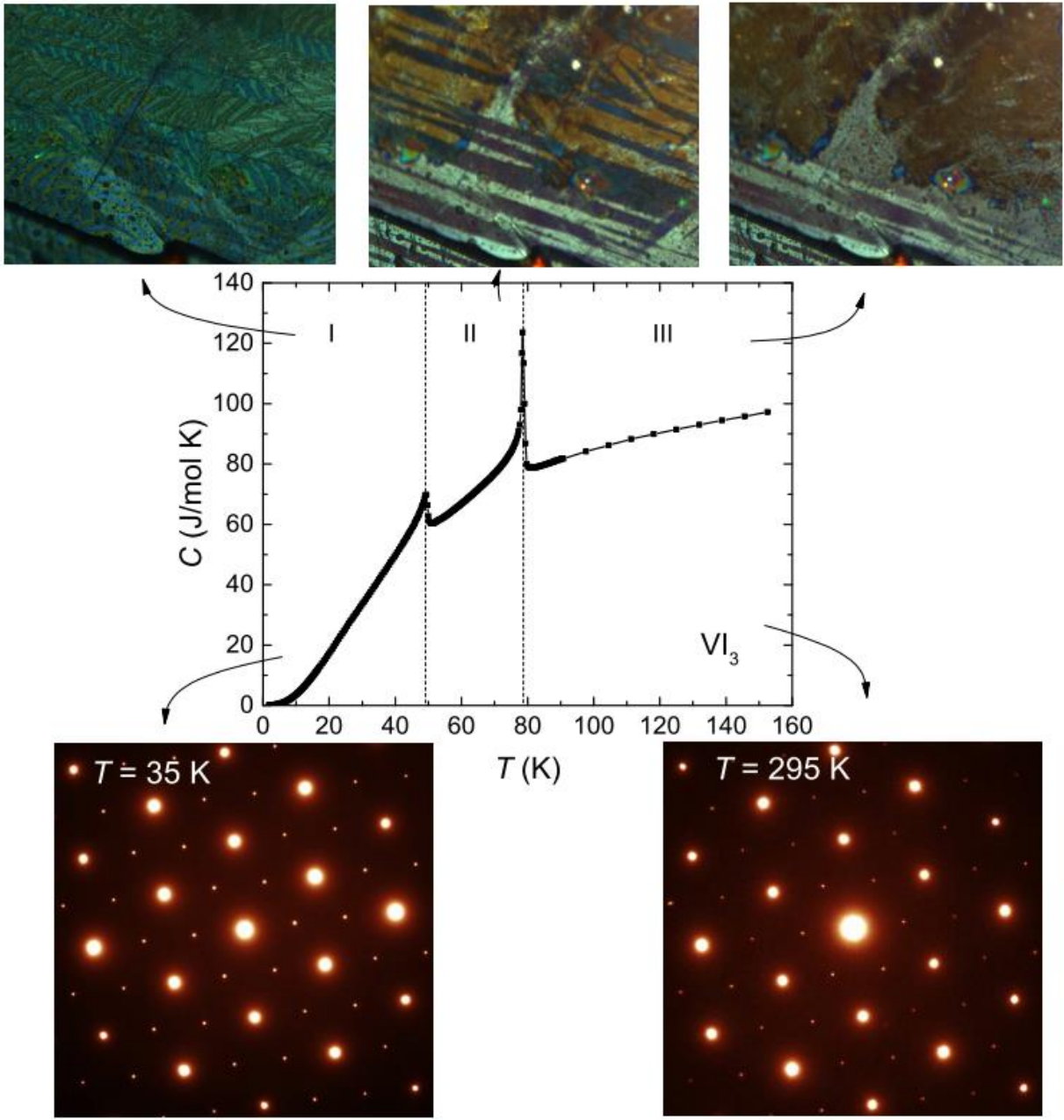

**Table 1.** Wyckoff positions, coordinates, occupancies, and equivalent isotropic displacement parameters respectively for $V_{1.04(1)}I_3$ single-crystal measured at 100(1) K. $U_{eq}$ is one third of the trace of the orthogonalized $U_{ij}$ tensor.

| Atom | Wyck. Site | x | y | z | Occupancy | $U_{eq}/U_{iso}$ |
|---|---|---|---|---|---|---|
| I1 | 18f | 0.34897(3) | 0.34802(3) | 0.08010(2) | 1 | 0.00666(6) |
| V1 | 6c | 0 | 0 | 0.33366(5) | 1 | 0.00684(15) |
| V2 | 3a | 0 | 0 | 0 | 0.04(1) | 0.012(5) |

# Supporting Information

**VI$_3$: a new layered ferromagnetic semiconductor**

*Tai Kong[1], Karoline Stolze[1], Erik Timmons[2], Jing Tao[3], Danrui Ni[1], Shu Guo[1], Ruslan Prozorov[2], Robert J. Cava[1]*

**Table S1.** Crystallographic data and details of the structure determination of V$_{1.04(1)}$I$_3$ derived from single-crystal experiments measured at 100(1) K.

| | |
|---|---|
| Sum Formula | V$_{1.04(1)}$I$_3$ |
| Formula weight / (g · mol$^{-1}$) | 433.85 |
| Crystal System | trigonal |
| Space group | *R*–3 (No. 148) |
| Formula units per cell, *Z* | 6 |
| Lattice parameter *a* / Å | 6.8879(3) |
| *c* / Å | 19.8139(9) |
| Cell volume / (Å$^3$) | 814.09(8) |
| Calculated density / (g · cm$^{-3}$) | 5.310 |
| Radiation | $\lambda$(Mo-*Kα*) = 0.71073 Å |
| | $2\theta \leq 72.58°$ |
| Data range | $-11 \leq h \leq 11$ |
| | $-11 \leq k \leq 11$ |
| | $-33 \leq l \leq 33$ |
| Absorption coefficient / mm$^{-1}$ | 18.78 |
| Measured reflections | 17867 |
| Independent reflections | 875 |
| Reflections with $I > 2\sigma(I)$ | 803 |
| *R*(int) | 0.044 |
| *R*($\sigma$) | 0.015 |
| No. of parameters | 15 |
| $R_1$(obs) | 0.025 |
| $R_1$(all $F_o$) | 0.028 |
| $wR_2$(all $F_o$) | 0.046 |
| Residual electron density / (e · Å$^{-3}$) | 1.86 to –1.56 |

**Table S2.** Anisotropic displacement parameters for $V_{1.04(1)}I_3$ single-crystal measured at 100(1) K. The coefficients $U_{ij}$ (/Å$^2$) of the tensor of the anisotropic temperature factor of atoms are defined by $\exp\{-2\pi^2[U_{11}h^2a^{*2} + \cdots + 2U_{23}klb^*c^*]\}$.

| Atom | $U_{11}$ | $U_{22}$ | $U_{33}$ | $U_{12}$ | $U_{13}$ | $U_{23}$ |
|---|---|---|---|---|---|---|
| I1 | 0.00779(9) | 0.00772(9) | 0.00600(8) | 0.00502(7) | 0.00136(6) | 0.00130(6) |
| V1 | 0.0065(2) | 0.0065(2) | 0.0074(4) | 0.00327(11) | 0 | 0 |

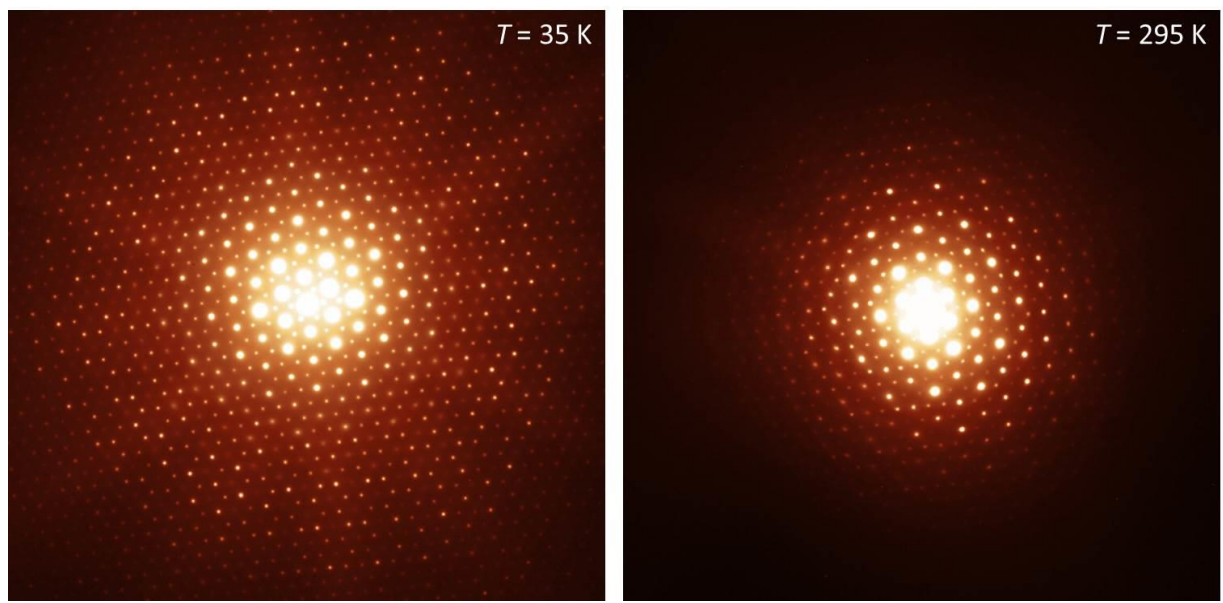

Figure S1. Electron diffraction patterns at 295 K and 35 K showing a larger amount of the reciprocal lattice.